\documentclass[
  ,final            
  ,numberedheadings 
]{aipproc}

\layoutstyle{6x9}

\begin{document}

\author{A.\ Lobo}{
  address={Instituto de Ciencias del Espacio (ICE), Barcelona, Spain},
  altaddress={Institut d'Estudis Espacials de Catalunya (IEEC),
  Barcelona, Spain}}
\author{M.\ Nofrarias}{
  address={Institut d'Estudis Espacials de Catalunya (IEEC),
  Barcelona, Spain}}
\author{J.\ Ramos--Castro}{
  address={Universitat Polit\`ecnica de Catalunya (UPC),
  Barcelona, Spain}}
\author{J.\ Sanjuan}{
  address={Institut d'Estudis Espacials de Catalunya (IEEC),
  Barcelona, Spain}}
\author{A.\ Conchillo}{
  address={Institut d'Estudis Espacials de Catalunya (IEEC),
  Barcelona, Spain}}
\author{J.A.\ Ortega}{
  address={Institut d'Estudis Espacials de Catalunya (IEEC),
  Barcelona, Spain}}
\author{X.\ Xirgu}{
  address={Institut d'Estudis Espacials de Catalunya (IEEC),
  Barcelona, Spain}}
\author{H.\ Araujo}{
  address={Imperial College of London (UK)}}
\author{C.\ Boatella}{
  address={Institut d'Estudis Espacials de Catalunya (IEEC),
  Barcelona, Spain}}
\author{M.\ Chmeissani}{
  address={Institut de F\'\i sica d'Altes Energies (IFAE),
  Barcelona, Spain}}
\author{C.\ Grimani}{
  address={Universit\`a di Urbino and INFN (Italy)}}
\author{C.\ Puigdengoles}{
  address={Institut de F\'\i sica d'Altes Energies (IFAE),
  Barcelona, Spain}}
\author{P.\ Wass}{
  address={Imperial College of London (UK)}}
\author{E.\ Garc\'\i a-Berro}{
  address={Universitat Polit\`ecnica de Catalunya (UPC),
  Barcelona, Spain}}
\author{S.\ Garc\'\i a}{
  address={At\'\i pic, Parc Tecnol\`ogic del Vall\'es,
  Barcelona, Spain}}
\author{L.M.\ Mart\'\i nez}{
  address={At\'\i pic, Parc Tecnol\`ogic del Vall\'es,
  Barcelona, Spain}}
\author{G.\ Montero}{
  address={At\'\i pic, Parc Tecnol\`ogic del Vall\'es,
  Barcelona, Spain}}

\title{In-flight Diagnostics in \textsl{LISA Pathfinder}}


\begin{abstract}
LISA PathFinder (LPF) will be flown with the objective to test in space
key technologies for LISA. However its sensitivity goals are, for good
reason, one order of magnitude less than those which LISA will have to
meet, both in drag-free and optical metrology requirements, and in the
observation frequency band. While the expected success of LPF will of
course be of itself a major step forward to LISA, one might not forget
that a further improvement by an order of magnitude in performance will
still be needed. Clues for the last leap are to be derived from proper
disentanglement of the various sources of noise which contribute to
the total noise, as measured in flight during the PathFinder mission.
This paper describes the principles, workings and requirements of one
of the key tools to serve the above objective: the diagnostics subsystem.
This consists in sets of temperature, magnetic field, and particle counter
sensors, together with generators of controlled thermal and magnetic
perturbations. At least during the commissioning phase, the latter will
be used to identify \emph{feed-through} coefficients between diagnostics
sensor readings and associated actual noise contributions. A brief
progress report of the current state of development of the diagnostics
subsystem will be given as well.
\end{abstract}

\maketitle

\section{Introduction}
\label{Lobo_sec-1}

LISA (Laser Interferometer Space Antenna), the joint ESA--NASA mission
to place a Gravitational Wave (GW) detector in heliocentric orbit, is
scheduled to fly within the next decade. The main objective of LISA is
to observe GWs in a frequency band around 1 mHz, where many interesting
sources are expected, but where earth based antennas are (by far) not
sensitive to: many galactic binaries, massive black holes in distant
galaxies and (perhaps) primeval GWs are amongst the signals LISA is
expected to sight, at a minimum. LISA will measure ambient GW induced phase
shifts in beams of laser light bouncing back and forth between freely
falling test masses. According to basic theoretical principles (see
e.g.~\cite{Lobo_lobo92}), this requires the nominal distance between the
test masses to be of the order of $c/4\nu$, where $\nu$ is the
frequency of the incoming GW and $c\/$ is the speed of light. For
$\nu$\,$\sim$\,1~mHz this gives an arm length of a few million~km.
For LISA, 5$\times$10$^6$ km has been baselined, and the mission is
defined as a formation of three spacecraft in a triangular configuration,
5$\times$10$^6$~km to the side~\cite{Lobo_lisa00}. For this, a heliocentric
orbit, 1~AU from the Sun, is foreseen. 

The key concept for LISA to work is to ensure that the test masses are
in nominal \emph{free fall}, i.e., that they follow the geodesics of the
local gravitational field. GWs of a given frequency then show up in the
detector as \emph{differential accelerations} between the test masses,
at that frequency. This is why the sensitivity requirement for LISA is
commonly stated in terms of relative acceleration noise. The current
baseline is
\begin{equation}
  S^{1/2}_{\Delta a}(\nu) = 3\times 10^{-15}\,\left[1 + \left(
  \frac{\nu}{3\ {\rm mHz}}\right)^2\right]\,
  {\rm m\,s}^{-2}\,{\rm Hz}^{\!-1/2}\ ,\quad
  10^{-4}\,{\rm Hz}\,\leq\,\nu\,\leq\,10^{-1}\,{\rm Hz}
  \label{Lobo_eq.1}
\end{equation}

This is a rather formidable requirement: picometre interferometry will
be needed, and an extremely quiet environment for the test masses must
be maintained. The latter is provided by a so called \emph{drag-free}
subsystem, which consists in a high precision test mass position sensing
device, called \textsl{Gravitational Reference Sensor} (GRS), working in
combination with a set of micro-thrusters which keep the spacecraft in
orbit \emph{following} the test masses.

The European Space Agency (\textsl{ESA}) has decided to launch a previous
technology demonstrator to check in flight the feasibility issues of LISA.
The mission is called LISA PathFinder (LPF), and is scheduled fly in 2009.
Seven European countries participate in this mission.

The payload on board LPF is the \textsl{LISA Technolgy Package} (LTP),
and includes several subsystems and interfaces, both amongst them and
with the space platform itself, whose rigorous control is part of the
experiment. The main purpose of the LTP is to test in space the key
technologies for LISA. The concept is to use only two freely floating
masses, and to follow their evolutions in a reduced size configuration:
a LISA arm is squeezed from 5 million kilometres to about 30 centimetres,
and a single spacecraft hosts both test masses, where drag-free and
interferometry is implemented. In addition, a relaxed sensitivity
requirement is adopted for LPF:
\begin{equation}
  S^{1/2}_{\Delta a}(\nu) = 3\times 10^{-14}\,\left[1 + \left(
  \frac{\nu}{3\ {\rm mHz}}\right)^2\right]\,
  {\rm m\,s}^{-2}\,{\rm Hz}^{\!-1/2}\ ,\quad
  1\,{\rm mHz}\,\leq\,\nu\,\leq\,30\,{\rm mHz}
  \label{Lobo_eq.2}
\end{equation}
which is an order of magnitude below that of LISA, both in spectral
density magnitude and in bandwidth.

One of the subsystems of the LTP is the so called \textsl{Data and
Diagnostics Subsystem} (DDS), which consists in a series of items
intended to monitor various factors of disturbance inside the payload.
The purpose of these instruments is to provide information to split up
the total system readout noise into different components, with the
the goal of both diagnosing LTP performance, and guiding the search
for the final sensitivity leap from equation~\eqref{Lobo_eq.2}
to~\eqref{Lobo_eq.1}. In addition, the DDS also provides the \emph{Data
Management Unit} (DMU), with several control and feedback functions,
and on-board data analysis duties.

In the following pages we briefly review the motivation, significance
and current state of development of the LTP DDS.

\section{Diagnostics elements}
\label{Lobo_sec-2}

Three types of disturbances have been identified which need to be
diagnosed in the LTP:

\begin{itemize}
\setlength{\itemsep}{-0.2 ex}   
 \item Temperature fluctuations
 \item Magnetic fluctuations
 \item Incident fluxes of charged particles
\end{itemize}

A quantitative assessment of the actual contribution of each of these
items to the total system noise requires not only to measure them but
also knowledge of \emph{transfer functions}. The latter provide the
relationship between e.g. a temperature change and the associated
system readout ---normally in frequency domain. Such transfer functions
are determined on the basis of in-flight experiments, which consist in
measuring the effect of artificially induced perturbations on the system
response. Controlled perturbations are applied by means of \emph{heaters}
and \emph{magnetic coils} at suitable locations and with suitable properties.
These are also part of the DDS.

\subsection{Noise debugging philosophy}
\label{Lobo_sec-2.0}

Let $\alpha$ be a controlled disturbance applied to the system. In practice,
$\alpha$ will be a thermal gradient or a magnetic field and magnetic field
gradient. It is expedient that $\alpha$ be a \emph{coherent} signal, as
this will make its identification in the readout data stream easier. Let
$y(t;\alpha)$ be the instrumental response data stream ---we leave it
generic, as it can be phasemeter data, force on the test masses, or some
other suitable magnitude. Using $y(t;\alpha)$, we calculate the
\emph{feedthrough} coefficient
\begin{equation}
  F = \frac{\partial y}{\partial\alpha}
  \label{Lobo_eq.2a}
\end{equation}

Normally, $\alpha$ will be strong enough that it can be unambiguously
detected in the output data, $y(t;\alpha)$. Actually, the requirement
is that it be seen with a SNR of 50~\cite{Lobo_sv05}. The idea is to
extrapolate the value of the \emph{feedthrough} coefficient $F\/$ to
the weaker disturbance regime prevailing during science mode mission
operation, so that ``$\alpha$--meter'' readings (i.e., thermometers and
magnetometers readings) can be translated into $y\/$-noise by multiplication
by $F$.

With such operation, we can evaluate the contributions of magnetic and
temperature fluctuations noise to the total LTP noise budget. In addition,
we know the sources of those contributions ---since they are provided by
the magnetometers' and thermometers' readings. This is essential information
to determine the line of improvement of system design in view to improve
the LTP sensitivity towards the more demanding LISA goal,
eq.~\eqref{Lobo_eq.1}. This is the reason why the LTP Diagnostics is
such a necessary subsystem: it would surely be less relevant should
LPF be the final mission, i.e., with no further projection into LISA.

The just described schematics is basically conceptual. However, its
practical implementation has a number of complications which need to
be thoroughly worked on to make it useful. For example, there are many
thermometers measuring the effect of several heaters. Hence we typically
have a \emph{multivariate problem}, with all its added nuances and
technical difficulties ---see Miquel Nofrarias \emph{et al.}, also
in this volume.

\subsection{Thermal diagnostics}
\label{Lobo_sec-2.1}

Thermal gradients are a major source of concern since they affect almost
every component of the LTP. In the
GRS, where the test masses are placed inside a vacuum enclosure,
\emph{radiation pressure} and \emph{radiometer effects} have been
identified as the major sources of temperature fluctuation noise
---see~\cite{Lobo_us06} for a comprehensive discussion. These can be
quantified and reliably modeled. The optical elements in the Optical
Metrology System (OMS) are also affected by random temperature gradients,
but their impact on the readout is much more elusive to detailed modeling
in this case.

The top level Science Requirements~\cite{Lobo_sv05} establish that temperature
fluctuations noise should account for 10\,\% of the total instrument
noise budget, at most. This sets a limit on acceptable temperature
fluctuation noise at~\cite{Lobo_us06}
\begin{equation}
  S^{1/2}_{T}(\nu) \leq 10^{-4}\ {\rm K\,Hz}^{-1/2}\ ,\qquad
  1\,{\rm mHz}\,\leq\,\nu\,\leq\,30\,{\rm mHz}
  \label{Lobo_eq.3}
\end{equation}

\begin{figure}[b]
\includegraphics[width=0.75\textwidth]{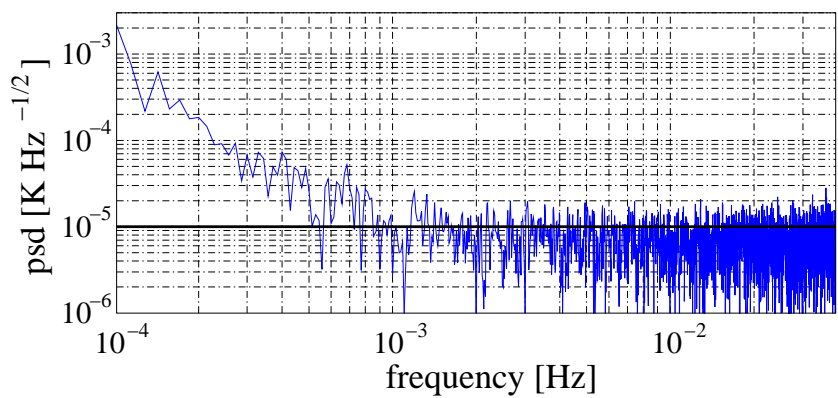}
\caption{Power spectral density of thermistors plus their
front-end electronics. Note that it is in the required level
of~10$^{-5}$\,K\,Hz$^{-1/2}$ throughout the MBW.
\label{Lobo_fig.1}}
\end{figure}

The satellite design must of course comply with this limit. However,
temperature measurements need to be taken at various strategic spots
to gauge the actual temperature environment conditions. Resolution of
such measurements must be more exigent, and 10$^{-5}$\,K\,Hz$^{-1/2}$
is required for them, within the LTP measuring bandwidth (MBW).
A~total of 22 thermometers will be distributed
across the LTP, close to the test masses inside the GRS, in the optical
bench and in the LCA (LTP Core Assembly) mounting struts\footnote{
They are carbon fibre beams with titanium braces which keep the LCA tied
to the spacecraft structure.}. The sensors have been chosen to be
\emph{thermistors}; the associated electronics has been designed, and
prototypes built and tested~\cite{Lobo_ddd}. They perform to full satisfaction
---see Figure~\ref{Lobo_fig.1}---, and are currently being submitted to
\emph{baking tests}. The reason for such tests is that the entire GRS
structure, including of course temperature sensors and wirings, will be
heated up in order to inhibit outgassing inside it.

But, once the temperature readings are given, next question is: how do
we extract useful information from them? We clearly need to know the
relationship between temperature and system readout. As mentioned in the
previous section, thermal control signals must be applied to the LTP,
then measure its response. Details on this matter will be found in
Miquel Nofrarias \emph{et al.}, also in this volume.

\subsection{Magnetic fluctuations}
\label{Lobo_sec-2.2}

Each of the LTP test masses is a cube, 46 mm to the side, 1.96 kg of mass,
and made of an alloy of 70\,\% gold and 30\,\% platinum. This has a very
low magnetic susceptibility, $|\chi|$\,$\leq$\,10$^{-5}$, and a very low
remnant magnetic moment, too: $|{\bf m}_0|$\,$\leq$\,10$^{-8}$\,A\,m$^2$.
Nevertheless, fluctuations of these magnitudes, as well as of environmental
magnetic fields and gradients, will contribute to the system's overall
noise. Just like temperature fluctuations, in-flight measurements of
magnetic disturbances are necessary for a thorough assessment of mission
design, i.e., how much \emph{magnetic noise} is present in the readout.

Magnetic diagnostics pose significant measurement problems. Indeed,
sensitive magnetometers make use of high susceptibility magnetic core
materials, which may therefore not get too close to the TMs, hence field
value extrapolations are required to estimate their value at the TMs.
Algorithms for this are currently under investigation~\cite{Lobo_atipic},
and are at present based on \emph{offline} statistical procedures.
Error margins are in the order of 10\,\% to 40\,\%, which is quite
good for the difficult magnetometer configuration which has been
baselined at system level: four \emph{fluxgate} magnetometers at
distances in excess of 15~cm to the closest TM. Further refinements
are being worked on, and improvements are shortly expected.

Magnetic fields have a distinctive feature: this is the fact that, so
long as the TM susceptibility is non-zero, magnetisation on the TMs is
induced, hence a force on them. If the magnetic field is {\bf B} then
the force is given by
\begin{equation}
 {\bf F} = \left\langle\left[\left(
 {\bf m}_0 + \frac{\chi V}{\mu_0}\,{\bf B}\right)\cdot\nabla\right]
 {\bf B}\right\rangle
 \label{Lobo_eq.4}
\end{equation}
where $V\/$ is the test mass volume, and $\langle-\rangle$ stands for
volume average within the test mass. Equation~\eqref{Lobo_eq.4} shows that
magnetic field intensity relates \emph{non-linearly} to the associated
force, and this has two important consequences: first, magnetic field
and gradient \emph{DC} components need to be properly monitored, as
they couple to each other's fluctuating components; and second, high
frequency magnetic field fluctuations result in \emph{DC} ---or low
frequency--- forces because of the quadratic dependence of {\bf F}
on {\bf B}.

Magnetic diagnostics also include the concept of controlled generation
of magnetic fields ---see section~\ref{Lobo_sec-2.0} above. This
is provided by magnetic induction coils (one per TM) which produce
\emph{non-homogeneous} magnetic fields in the vicinity of the TMs. The
quadratic nature of the magnetic force shown by equation~\eqref{Lobo_eq.4}
results in a two-frequency response to a one-frequency input:
a sinusoidal signal of frequency $\omega$ in the coil generates
a response at the same frequency which is proportional to the TM
remnant magnetic moment ${\bf m}_0$, plus a signal at 2$\omega$,
proportional to the TM susceptibility, $\chi$. This means that
${\bf m}_0$ and $\chi$ can both be re-measured in flight, and that
magnetic noise debugging can proceed thereafter, as explained above.

\subsection{Radiation monitor}

Cosmic rays and certain solar events contain ionising particles which will
hit the LTP in flight, thus causing spurious signals in the GRS. These
particles are mostly protons, with 10\,\% or less of He nuclei, and a
tiny fraction of heavier nuclei, electrons and solar ions. Charging rates
and the properties of noise caused by charging vary depending on whether
the particle flux comes from Galactic Cosmic Rays (GCR) or is augmented
by Solar Energetic Particles (SEP). The reason is that the two types
of radiation present different energy spectra. Although average charging
rates are detected by a dedicated measurement provided by the GRS, temporal
\emph{fluctuations} of the GCR flux and SEP can contaminate the data.
A particle counter is thus necessary to provide correlations between the
flux of energetic particles and the instantaneous charging rates observed
in the test masses. In addition, the device must have the ability to
distinguish SEP events from GCRs, and this consequently means it needs to
determine the \emph{energy spectra} of the detected particles. Finally, not
all charged particles hitting the satellite structure will make it to the
test masses, as that structure itself has a certain \emph{stopping power}.
The particle counter must only be triggered by those particles having
enough energy to reach the TMs, hence it must be properly \emph{shielded}.
Simulation work indicates that only ions with energies larger than
$\sim$100\,MeV should be counted~\cite{Lobo_MeV}. The particle counter
together with the above added capabilities is known as Radiation Monitor
(RM). Contrary to the previous diagnostics, the RM does not require
in-flight calibration.

An RM prototype has been designed and built in IFAE (Barcelona), with
essential collaboration with Imperial College (London) and the University
of Urbino~\cite{Lobo_ifae}. The design concept consists in a pair of PIN
diodes in telescopic configuration; incoming ionising particles generate
charge in each PIN diode which is measured by dedicated electronics. This
charge deposit provides a measurement of the incoming particle's energy,
but it cannot tell e.g.\ protons from photons. Coincident events in both
PIN diodes exclude photons, so spectral analysis is done on coincidence
data to select only charged particle impacts, and to resolve SEP from
GCR events. The significance of RM data is not based on individual
events but on statistics of longer data stretches.

A most important part of the RM is the \emph{shielding} which protects
the PIN diodes against impacts of particles with energies below 100~MeV.
This has a cubic copper profile, with rounded vertices, some 6~mm thick.
The RM has been thoroughly tested to ensure its workings under electronically
generated impulses. It has also been submitted to laboratory proton beam
irradiation at the Paul Scherrer Institute (PSI) in Switzerland, and the
results are very satisfactory in general ---see P.\ Wass \emph{et al.},
also in this volume.

A concern raised by the PSI test was the possible activation of the copper
shield under excess irradiation conditions. These are considered unlikely
in Lagrange-L1, but alternatives will also be considered and analysed during
the following test, with fully space qualified materials.

\section{The Data Management Unit (DMU)}

The DMU is the LTP computer. It holds full control of the diagnostics
elements, i.e., powers them on and off (according to function programmes),
acquires data, and processes them. Additionally, a number of other LTP
functions are also managed by the DMU. The DMU is however subordinated
to the mission on-board computer (OBC).

The hardware of the DMU consists in three major electronic boards: the
Power Distribution Unit (PDU), Data Acquisition Unit (DAU), and Data
Processing Unit (DPU). However, each of these boards is \emph{duplicated}
for \emph{redundancy}. The PDUs and the DPUs work in a hot-cold redundancy
scheme, which means only one of these boards is on at any given time. The
DAUs are instead both operative at all times, and tasks are strategically
distributed between them to minimise losses in case one of them fails.

The software running in this machine has two major components: the
\emph{Boot Software} (BSW) and the \emph{Application Software} (ASW).
Writing of both pieces of software requires intensive interaction with
several other mission partners, as a significant part of it interfaces
with subsystems different from the DDS. Progress is good so far, and
the developers team includes the industrial contractor personnel (NTE).
Further details will be found in J.A.\ Ortega \emph{at al.} report,
also in this volume.

\section{Conclusion}

Shortly after the end of the LISA Symposium, the DDS went through a
Delta-PDR (Preliminary Design Review) in July-2006, in which two items
pending from the initial PDR (September 2005) were reviewed again.
PDR will be closed after a number of actions are completed. Next
landmark is the Critical Design Review (CDR), in a few months time.
Hardware procurement and assembly will however start even before, as
there are a number of items in which no changes can be reasonably
expected. We are confident that conditions look good for a successful
completion of the DDS, a very important mission subsystem, of course
framed within a global mission progress~success.

\textbf{\textsf{\small Acknowledgement:}} We thank Ministerio de
Educaci\'on y Ciencia for support under contract ESP2004-01647.

\end{document}